\documentclass[12pt]{iopart}
\begin{document}
\newcommand{\eqref}[1]{(\ref{#1})}
\newcommand{\Eq}[1]{Eq.~(\ref{#1})}
\newcommand{\Eqs}[1]{Eqs.~(\ref{#1})}
\newcommand{\Ref}[1]{Ref.~\cite{#1}}
\newcommand{\Refs}[1]{Refs.~\cite{#1}}
\newcommand{\Sec}[1]{Sec.~\ref{#1}}

\title[A regularized Lagrange-mesh method]{A regularized Lagrange-mesh method based on an orthonormal Lagrange-Laguerre basis}
\date{}
\author{J\'er\'emy Dohet-Eraly}
\address{TRIUMF, 4004 Wesbrook Mall, Vancouver BC V6T 2A3, Canada.\\
Istituto Nazionale di Fisica Nucleare, Sezione di Pisa, Largo B.
Pontecorvo 3, I-56127 Pisa, Italy.}
\ead{dohet@pi.infn.it}
\begin{abstract}
The Lagrange-mesh method is an approximate variational approach having the form of a mesh calculation because of the use of a Gauss quadrature. Although this method provides accurate results in many problems with small number of mesh points, its accuracy can be strongly reduced by the presence of singularities in the potential term.
In this paper, a new regularized Lagrange-Laguerre mesh, based on \textit{exactly} orthonormal Lagrange functions, is devised. 
It is applied to two solvable radial potentials: the harmonic-oscillator and Coulomb potentials. In spite of the singularities of the Coulomb and centrifugal potentials, accurate bound-state energies are obtained for all partial waves. 
The analysis of these results and a comparison with other Lagrange-mesh calculations lead to a simple rule to predict in which cases a singularity does induce or not a significant loss of accuracy in Lagrange-mesh calculations.
In addition, the Lagrange-Laguerre-mesh approach is applied to the evaluation of phase shifts via integral relations. Small numbers of mesh points suffice to provide very accurate results.
\end{abstract}

\submitto{\jpa}
\maketitle
\section{Introduction}
The Lagrange-mesh method~\cite{BH86,Ba15} is a simple and often accurate method for solving quantum-mechanical problems. It can be seen as an approximate variational calculation using a basis of Lagrange functions associated with a mesh and the Gauss quadrature associated with this mesh to evaluate most of the matrix elements. For reasons not perfectly understood~\cite{BHV02}, the Lagrange-mesh method is in many applications as accurate as the corresponding variational approach. However, the success of the Lagrange-mesh method relies ultimately on the validity of the Gauss quadrature. Therefore, when the quality of the Gauss-quadrature approximation is poor, the Lagrange-mesh method is inaccurate. In particular, it happens when the potential is singular but surprisingly, only for the lowest partial waves~\cite{BHV02}.
For some problems, the high accuracy of the Lagrange-mesh method can be restored by means of a regularization technique~\cite{VMB93}. For other problems, such a regularization technique is not available, yet. It is the case for instance for the study of a Coulomb system with more than three particles.  In the case of three particles, the presence of singularities in the potential restricts the high quality of the Lagrange-mesh method to some specific systems of coordinates~\cite{He02}. A better understanding of the regularization process, its effects and its necessity, could help to develop new regularization techniques and thus, broaden the scope of applicability of the Lagrange-mesh method.

In this paper, a regularized Lagrange mesh based on orthonormal Lagrange-Laguerre functions is developed and applied to bound-state calculations. A comparison with other Lagrange meshes provides interesting insight into the regularization process. The Lagrange-mesh method is also extended to the phase-shift calculations by evaluating integral relations derived from the Kohn variational principle and adapted from~\cite{KVB10}. The integral relations are computed with the Gauss quadrature associated with the Lagrange mesh, which make their evaluation particularly easy. The accuracy of the approach is analyzed.

In \Sec{S2}, the application of the Lagrange-mesh method to the study of bound states and to the calculation of phase shifts is outlined. The Lagrange meshes considered in this work are defined in \Sec{S3}. Formulae associated with these meshes and newly derived are given. In \Sec{S4}, the Lagrange-mesh method is applied to different test cases and the obtained results are discussed. Concluding remarks are given in \Sec{S5}.
Finally, in the appendix, the Lagrange-mesh method is applied to a somehow exotic but illuminating problem: the study of bound states in a two dimensional space.

\section{Model}\label{S2}
For the partial wave of orbital angular momentum $\ell$, the radial Schr\"odinger equation for a particle with mass $M$ in a central potential $V(r)$  reads
\begin{equation}\label{Sch}
H_\ell u_{k\ell}(r) =
\left[-\frac{1}{2}\frac{\rmd^2}{\rmd r^2} +\frac{\ell (\ell+1)}{2 r^2}+V(r)\right]u_{k\ell}(r)=E u_{k\ell}(r)
\end{equation}
with the boundary condition
\begin{equation}
u_{k\ell}(0) =0,
\end{equation}
where $E$ is the energy of the particle, $k=\sqrt{2E}$ is the wave number, $r$ is the radius in spherical coordinates, and $\hbar=M=1$.
The potential $V(r)$ is assumed to have singularities, if any, only at the origin ($r=0$) or at infinity ($r=\infty$). Moreover, at the origin, the potential $V(r)$ is assumed to be less singular than the centrifugal term,
\begin{equation}
r^2 V(r)\mathop{\longrightarrow}\limits_{r\to 0} 0.
\end{equation}
With these hypotheses, the radial wave function behaves at the origin as
\begin{equation}\label{ori}
 u_{k\ell}(r) \mathop{\longrightarrow}\limits_{r\to 0} r^{l+1}.
\end{equation}
Let us first consider a purely discrete spectrum. 
Equation~\eqref{Sch} can be solved approximately by expanding the radial wave function $u_{k\ell}$ in some set of square-integrable functions $\{f_j\}_{j=1,\ldots,N}$,
\begin{equation}\label{uexp}
u_{k\ell}(r)=\sum^N_{j=1} c_j f_j(r).
\end{equation}
Then, the energies and the coefficients $c_j$ are determined variationally by solving the generalized eigenvalue problem
\begin{equation}\label{eigen}
\sum^N_{j=1} \langle f_i | H_\ell | f_j\rangle c_j =E \sum^N_{j=1} \langle f_i | f_j\rangle c_j 
\end{equation}
for $i=1,\ldots,N$. The eigenvalue problem is not generalized if the norm matrix is the identity, i.e., if the functions $f_j$ are orthonormal. 

Now, let us consider that the potential $V$ tends asymptotically to the Coulomb potential $V_C$,
\begin{equation}
r^2[V(r)-V_C(r)] \mathop{\longrightarrow}\limits_{r\to\infty} 0,
\end{equation}
where the Coulomb potential is given by $V_C(r)=Z/r$ with parameter $Z$. This potential has a possible discrete spectrum of negative energies and a continuum spectrum of positive energies. For any strictly positive energy, the Schr\"odinger equation~\eqref{Sch} has a non-square integrable solution with an asymptotic behavior given by
\begin{equation}\label{asymp}
u_{k\ell}(r) \mathop{\longrightarrow}\limits_{r\to\infty} A_{k\ell} [\cos\delta_\ell(E) F_\ell(\eta,kr)+\sin\delta_\ell(E) G_\ell(\eta,kr)],
\end{equation}
where $\eta=Z/k$, $A_{k\ell}$ is a normalization coefficient, $\delta_\ell(E)$ is the phase shift corresponding to partial wave $\ell$ and energy $E$, and $F_\ell$ and $G_\ell$ are the regular and irregular Coulomb functions.
For such a potential, the solutions of \eqref{eigen} with negative energies are again variational approximations of the bound-state wave functions while the solutions of \eqref{eigen} with positive energies correspond to pseudostates, which are square-integrable approximations of continuum states. Although these pseudostates do not have the proper asymptotic behavior \eqref{asymp} of a positive-energy wave function, they can be used to determine the phase shift via integral relations derived from the Kohn variational principle~\cite{KVB10},
\begin{equation}\label{KVP}
\tan\delta_\ell(E) =-\frac{ \int^\infty_0 [V(r)-V_C(r)] u_{k\ell}(r) F_\ell(\eta,k r) \rmd r}{
\int^\infty_0 [V(r)-V_C(r)] u_{k\ell}(r) \tilde{G}_\ell(\eta,k r) \rmd r+I_\gamma},
\end{equation}
where $\tilde{G}_\ell=G_\ell (1-\rme^{-\gamma r})^{l+1}$ with $\gamma>0$ is a regularization of the irregular Coulomb function $G_\ell$ and
\begin{eqnarray}
I_\gamma=\frac{1}{2} \int^\infty_0 u_{k\ell}(r) (1-\rme^{-\gamma r})^{l-1} (l+1) \gamma \rme^{-\gamma r}\nonumber\\
\times\left\{\gamma  \left[1-(l+1) \rme^{-\gamma r}\right]-2 (1-\rme^{-\gamma r})\frac{\rmd}{\rmd r}\right\}G_\ell(\eta,k r) \rmd r.
\end{eqnarray}\label{Ig}
Indeed, the integrals in \eqref{KVP} and \eqref{Ig} are short-ranged and therefore, only the internal part of the wave function has to be described accurately to provide  accurate phase shifts.
Relation~\eqref{KVP} is a straightforward generalization for an arbitrary angular momentum of relations given in \cite{KVB10}.
In theory, this relation is valid for any positive value of $\gamma$, including $\gamma=0$ with $I_\gamma=0$.
However, in practice, the radial wave function is not known exactly and some approximation of $u_{kl}$, in a finite basis for instance, like \eqref{uexp}, is used. 
In this case, relation \eqref{KVP} is only accurate for some range of $\gamma$ values. Several $\gamma$ values need to be tested to find a plateau where the sensitivity of the results to the $\gamma$ parameter is weak. This sensitivity study gives a valuable information about the numerical accuracy of the phase shift which is reached for a particular calculation.

In this work, the radial wave function $u_{k\ell}$ is expanded in a Lagrange basis~\cite{BH86,Ba15}. 
Such a basis is associated with a set of $N$ mesh points $\{r_i\}_{i=1,\ldots,N}$ and the Gauss quadrature associated with this mesh.
Each Lagrange function vanishes at all mesh points except one,
\begin{equation}\label{Lagcond}
f_j(r_i)=\lambda^{-1/2}_j \delta_{ij},
\end{equation}
where the constants $\lambda_j$ are the weights of the Gauss quadrature associated with the mesh,
\begin{equation}
\int^\infty_0 g(r) \rmd r \approx \sum^N_{k=1} \lambda_k g(r_k).
\end{equation}
The abscissae $r_k$ and the weights $\lambda_k$ of the Gauss quadrature are defined implicitly by the relations
\begin{equation}
\int^\infty_0 w(r) P_{2N-1}(r) \rmd r = \sum^N_{k=1} \lambda_k  w(r_k) P_{2N-1}(r_k),
\end{equation}
which are to be true for any polynomial $P_{2N-1}$ of degree lower than or equal to $2N-1$. The weight function $w$ defines the type of Gauss quadrature which is used. The choices of $w$ considered in this work are specified in the next section.

Direct consequences of property \eqref{Lagcond} are that, at the Gauss approximation, the Lagrange functions are orthonormal,
\begin{equation}
\langle f_i|f_j\rangle_G=\sum^N_{k=1} \lambda_k f_i(r_k) f_j(r_k)=\delta_{ij},
\end{equation}
and the potential matrix is diagonal,
\begin{equation}
\langle f_i|V|f_j\rangle_G=\sum^N_{k=1} \lambda_k f_i(r_k) V(r_k) f_j(r_k)=V(r_i) \delta_{ij},
\end{equation}
where $\delta_{ij}$ is the Kronecker delta and the subscript $G$ indicates that the Gauss quadrature is used.
Calculating the overlap and potential matrix elements by applying the Gauss quadrature makes their evaluation particularly simple and, in many cases, does not reduce significantly the accuracy of the method. However, in presence of \textit{singularities}, the application of the Gauss quadrature can reduce the accuracy of the results by several orders of magnitude~\cite{BHV02}. 
In this context, the term \textit{singularity} means that in the matrix element, the integrand divided by the weight function $w$ associated with the Gauss quadrature is singular. 
For instance, the centrifugal term ($\ell \ne 0$),
\begin{equation}
\langle f_i|\frac{\ell (\ell+1)}{2 r^2}|f_j\rangle=\frac{\ell (\ell+1)}{2} \int^\infty_0 \frac{f_i(r) f_i(r)}{r^2w(r)} w(r)\rmd r,
\end{equation}
is said \textit{singular} if the factor $f_i(r) f_i(r)/[r^2w(r)]$ is singular.
The presence or absence of \textit{singularities} depends on the operator and also on the considered Gauss quadrature or equivalently on the considered Lagrange-mesh. Studying in which cases the presence of a \textit{singularity} reduces the accuracy of the Lagrange-mesh method is one of the aims of this paper.
\section{Lagrange meshes}\label{S3}
Three Lagrange meshes are considered. All are based on a Gauss-Laguerre quadrature, i.e., a Gauss quadrature based on a weight function $w(r)=r^{\alpha} \rme^{-r}$ with $\alpha\ge 0$. Each mesh is made of $N$ Lagrange functions defined over $[0,\infty[$: the Lagrange-Laguerre functions~\cite{BH86,Ba15},
\begin{equation}\label{defnreg}
f^{(\alpha)}_j(r)=(-1)^j r^{1/2}_j (h^{(\alpha)}_N)^{-1/2} \frac{L^{(\alpha)}_N(r)}{r-r_j} r^{\alpha/2} \rme^{-r/2},
\end{equation}
the Lagrange-Laguerre functions regularized by $\sqrt{r}$,
\begin{equation}\label{defregsq}
\tilde{f}^{(\alpha)}_j(r) =\sqrt{r/r_j}  f^{(\alpha)}_j(r),
\end{equation}
and the Lagrange-Laguerre functions regularized by $r$~\cite{VMB93,Ba15},
\begin{equation}\label{defreg}
\hat{f}^{(\alpha)}_j(r) = \left(r/r_j\right)  f^{(\alpha)}_j(r)
\end{equation}
with $j=1,\ldots ,N$. The mesh points $r_j$ are the zeros of the generalized Laguerre polynomial $L^{(\alpha)}_N$. The normalization coefficients $h^{(\alpha)}_N$ are given by
\begin{equation}
h^{(\alpha)}_N=\Gamma(N+\alpha+1)/N!.
\end{equation}
All Lagrange functions are zero at each mesh point but one,
\begin{equation}\label{fjxi}
f^{(\alpha)}_j(r_i)=\tilde{f}^{(\alpha)}_j(r_i)=\hat{f}^{(\alpha)}_j(r_i)=\lambda^{-1/2}_j \delta_{ij},
\end{equation}
where the $\lambda_j$ are the Gauss-Laguerre weights. They are defined such that the Gauss quadrature
\begin{equation}
\int^\infty_0 g(r) \rmd r \approx \sum^N_{i=1} \lambda_i g(r_i)
\end{equation}
is exact if $g$ is any polynomial of degree lower than or equal to $2N-1$ times $r^{\alpha} \rme^{-r}$. 
While not explicitly denoted, the Gauss weights $\lambda_j$ depend on $N$ and $\alpha$.

Note that the set of basis functions $f^{(\alpha)}_j$, $\tilde{f}^{(\alpha-1)}_j$, and $\hat{f}^{(\alpha-2)}_j$ define the same vector space. 
However, when the Gauss quadrature is used, the results obtained with these three meshes differ, sometimes significantly.
While Lagrange meshes based on $f^{(\alpha)}_j$ and $\hat{f}^{(\alpha)}_j$ Lagrange functions have widely been used for solving radial Schr\"odinger equations \cite{BH86,VMB93,BHV02,Ba15}, the interest of Lagrange mesh regularized by $\sqrt{r}$, $\{\tilde{f}^{(\alpha)}_j\}$, has been ignored. Therefore, I mainly focus on this mesh in the rest of this section.

The exactness of the Gauss quadrature for the overlap matrix elements of the $\tilde{f}^{(\alpha)}_j$ functions combined with relation~\eqref{fjxi} shows that $\{\tilde{f}^{(\alpha)}_j\}$ is a set of orthonormal functions,
\begin{equation}\label{1}
\langle \tilde{f}^{(\alpha)}_i |\tilde{f}^{(\alpha)}_j\rangle=\langle \tilde{f}^{(\alpha)}_i |\tilde{f}^{(\alpha)}_j\rangle_G=\delta_{ij},
\end{equation}
like the set of non-regularized functions $\{f^{(\alpha)}_j\}$ but contrary to the set of regularized functions $\{\hat{f}^{(\alpha)}_j\}$.
The derivatives of the regularized functions $ \tilde{f}^{(\alpha)}_j$ evaluated at the mesh points can be deduced from the derivatives of the non-regularized function $f^{(\alpha)}_j$ at the same mesh points, which are given in \cite{Ba15}.
The first derivative reads, for $i\neq j$,
\begin{equation}\label{first}
\lambda^{1/2}_i \tilde{f}^{(\alpha) \prime}_j(r_i)=(-1)^{i-j} \frac{1}{r_i-r_j}
\end{equation}
and is zero for $i=j$,
\begin{equation}
\lambda^{1/2}_i \tilde{f}^{(\alpha) \prime}_i(r_i)=0.
\end{equation}
The second derivative reads, for $i\neq j$,
\begin{equation}
\lambda^{1/2}_i \tilde{f}^{(\alpha) \prime \prime}_j(r_i)=(-1)^{i-j+1} \frac{2}{(r_i-r_j)^2}
\end{equation}
and, for $i=j$,
\begin{equation}
\lambda^{1/2}_i \tilde{f}^{(\alpha) \prime \prime}_i(r_i)=-\frac{1}{12 r_i} \left[2(2N+\alpha+1)-r_i+\frac{1-\alpha^2}{r_i}\right].
\end{equation}

The matrix elements of $1/r$ and $\rmd/\rmd r$ between the Lagrange functions are exact at the Gauss quadrature because the integrands are a polynomial of degree lower than or equal to  $2N-1$ times $r^{\alpha} \rme^{-r}$. The matrix elements of $1/r$ are thus given by 
\begin{equation}
\langle \tilde{f}^{(\alpha)}_i |\frac{1}{r}  |\tilde{f}^{(\alpha)}_j\rangle=\langle \tilde{f}^{(\alpha)}_i|\frac{1}{r} |\tilde{f}^{(\alpha)}_j\rangle_G=\frac{1}{r_i} \delta_{ij}
\end{equation}
and the matrix elements of $\rmd/\rmd r$ by 
\begin{equation}
\langle \tilde{f}^{(\alpha)}_i |\frac{\rmd}{\rmd r}  |\tilde{f}^{(\alpha)}_j\rangle=\langle \tilde{f}^{(\alpha)}_i|\frac{\rmd}{\rmd r} \tilde{f}^{(\alpha)}_j\rangle_G= (-1)^{i-j}\frac{1}{r_i-r_j}
\end{equation}
for $i\neq j$ and
\begin{equation}
\langle \tilde{f}^{(\alpha)}_i |\frac{\rmd}{\rmd r}  |\tilde{f}^{(\alpha)}_i\rangle=\langle \tilde{f}^{(\alpha)}_i|\frac{\rmd}{\rmd r} \tilde{f}^{(\alpha)}_i\rangle_G=0
\end{equation}
for $i=j$.
The matrix element of $\rmd/\rmd r$ is not used in this work but is given for sake of completeness.
The matrix elements of $1/r^2$, $r$, and $r^2$ between the Lagrange functions are not exact at the Gauss quadrature but still have compact expressions,
\begin{equation}\label{1x2}
\langle \tilde{f}^{(\alpha)}_i |\frac{1}{r^2}  |\tilde{f}^{(\alpha)}_j\rangle=\frac{1}{r^2_i} \delta_{ij}+(-1)^{i-j}\frac{1}{\alpha r_i r_j},
\end{equation}
\begin{equation}\label{1x}
\langle \tilde{f}^{(\alpha)}_i | r |\tilde{f}^{(\alpha)}_j\rangle=r_i \delta_{ij}+(-1)^{i-j},
\end{equation}
and 
\begin{equation}\label{x2}
\langle \tilde{f}^{(\alpha)}_i |r^2 |\tilde{f}^{(\alpha)}_j\rangle=r^2_i \delta_{ij} +(-1)^{i-j} (2 N+\alpha+1+r_i+r_j).
\end{equation}
In the matrix element of $1/r^2$, $\alpha\neq 0$ is assumed to avoid a divergence.
In equations~\eqref{1x2}-\eqref{x2}, the first term of the r.h.s. corresponds to the matrix element calculated at the Gauss approximation.
Equations~\eqref{1x2} and \eqref{1x} can be deduced from the matrix elements of $1/r$ and $r^2$ between non-regularized functions $f^{(\alpha)}_j$, given in \cite{Ba15}. Relation~\eqref{x2} can be proven, by following an approach similar to the one proposed in Appendix 1 of \cite{VMB93}, based on the decomposition
\begin{equation}
r^2=(r-r_i)(r-r_j)+(r_i+r_j) r -r_i r_j.
\end{equation}
The integrals involving the last two terms are calculated from \eqref{1} and \eqref{1x}. The integral including the first term is easily calculated when the Lagrange functions $\tilde{f}^{(\alpha)}_i$ and $\tilde{f}^{(\alpha)}_j$ are expressed from \eqref{defnreg} and \eqref{defregsq} as functions of the generalized Laguerre polynomials $L^{(\alpha)}_N$.

The matrix element of $1/r^2$ involves an integrand which after division by the weight function $r^{\alpha} \rme^{-r}$ is singular. In spite of this fact, it is shown in the next section that using the Gauss quadrature for evaluating this matrix element still leads to accurate results.

The matrix elements of $\rmd^2/\rmd r^2$ between the Lagrange functions are not exact at the Gauss approximation except for $\alpha=1$. They are given, for $i\neq j$, by
\begin{equation}\label{dx2}
\langle \tilde{f}^{(\alpha)}_i |\frac{\rmd^2}{\rmd r^2}|\tilde{f}^{(\alpha)}_j\rangle=(-1)^{i-j+1} \left[ \frac{2}{(r_i-r_j)^2}+\frac{1-\alpha^2}{4 \alpha r_i r_j} \right]
\end{equation}
and, for $i=j$, by
\begin{equation}\label{last2}
\langle \tilde{f}^{(\alpha)}_i |\frac{\rmd^2}{\rmd r^2}|\tilde{f}^{(\alpha)}_i\rangle=\frac{-1}{12 r_i} \left[2(2N+\alpha+1)-r_i+\frac{1-\alpha^2}{r_i}
+\frac{3(1-\alpha^2)}{\alpha r_i}
\right],
\end{equation}
where $\alpha\neq 0$ is assumed to avoid a divergence.
The last term in the r.h.s\ of \eqref{dx2} and \eqref{last2} can be seen as a correcting term added to the Gauss approximation. This term is identically zero for $\alpha=1$.
Relations~\eqref{dx2} and \eqref{last2} can be proved by following an approach similar to the one proposed in Appendix 1 of \cite{VMB93}.
Relations~\eqref{first}-\eqref{x2}, \eqref{dx2}, and \eqref{last2} have been verified numerically. The corresponding formulae for the non-regularized and $r$-regularized meshes can be found in \cite{Ba15}.

For the sake of clarity, the unscaled versions of the Lagrange meshes have been presented in this section. 
However, to reach high accuracy with small numbers of mesh points, it is required to adapt the mesh to the size of the considered physical problem by an appropriate scaling, i.e., by considering the scaled Lagrange functions
\begin{equation}
{h^{-1/2} f^{(\alpha)}_j(r/h)}
\end{equation}
instead of the Lagrange functions $f^{(\alpha)}_j(r)$ and similarly for the other meshes. Parameter $h$ is called the scaling factor.
\section{Results}\label{S4}
First, the non-regularized Lagrange mesh \eqref{defnreg} with $\alpha=2$, the $\sqrt{r}$-regularized Lagrange mesh \eqref{defregsq} with $\alpha=1$, and the $r$-regularized Lagrange mesh \eqref{defreg} with $\alpha=0$ are applied to the study of bound states.
Since these Lagrange meshes correspond to the same vector space of basis functions, they are ideal for testing the impact of the Gauss quadrature on the accuracy of the different meshes. 
At the origin, all considered Lagrange functions behave like $r$.
They are thus able to reproduce the near-origin behavior \eqref{ori} of the radial wave function for any partial wave, provided $N>\ell$.

Two solvable potentials are considered, namely the harmonic-oscillator potential, $V(r)=r^2/2$, and the attractive Coulomb potential, $V(r)=-1/r$. For each one, the partial waves $s$, $p$, and $d$ are studied.
To study the accuracy of the Lagrange-mesh calculations in presence of singularities (centrifugal term and/or Coulomb), several calculations are performed: a variational calculation (referred as "var" in the tables) based on the $\sqrt{r}$-regularized Lagrange mesh, for which all matrix elements are calculated exactly by the formulae given in the previous section; two mesh calculations based on the regularized Lagrange functions $\tilde{f_j}$ and $\hat{f_j}$, referred, respectively, as "reg~$\sqrt{r}$" and "reg~$r$", for which all terms are calculated at the Gauss approximation; two mesh calculations based on the non-regularized Lagrange functions $f_j$: one (referred as "non reg") for which the Gauss approximation is made for the potential term and the centrifugal term and the other (referred as "non reg $V_G$") for which the Gauss approximation is only made for the potential term. 
The relative errors on the lowest-state energies for each considered partial wave ($s$, $p$, and $d$) for the harmonic-oscillator potential and for the Coulomb potential are given in Tables~\ref{tab1} and \ref{tab2}, respectively. 
They are defined by
\begin{equation}
\epsilon_{\rm rel}=\frac{E_{\rm app}-E_{\rm exact}}{|E_{\rm exact}|},
\end{equation}
where $E_{\rm app}$ and $E_{\rm exact}$ are the approximate and exact bound-state energies. With this definition, the variational principle holds if the relative error $\epsilon_{\rm rel}$ is positive.
The number of mesh points $N$ and the value of the scale parameter $h$ are given in the table captions. The value of the scale parameter is chosen to provide sensible results but is not really optimized since the aim is only to compare the accuracy of different approaches, not to get the most accurate solutions of the Sch\"odinger equations, which are analytically known anyway. To evaluate accurately the relative errors, all calculations have been performed with quadruple-precision arithmetic.
\begin{table}
\begin{center}
\begin{tabular}{rrrrrr}
\hline
$\ell$&var&reg $\sqrt{r}$&reg $r$&non reg&non reg $V_G$\\[0.1 cm]
\hline
0&1.9[-14]&6.8[-15]&9.4[-14]&1.4[-14]&1.4[-14]\\
1&4.4[-13]&4.6[-13]&-9.7[-14]&-2.8[-7]&4.4[-13]\\
2&2.7[-12]&-1.9[-13]&-9.1[-12]&2.5[-12]&2.5[-12]\\
\hline
\end{tabular}
\end{center}
\caption{Relative error $\epsilon_{\rm rel}$ on the lowest-state energy of the harmonic-oscillator potential $V(r)=r^2/2$ for the $s$, $p$, and $d$ waves obtained from several Lagrange meshes (see text for details) with $N=20$ and $h=0.09$. The notation $a[-b]$ stands for $a \times 10^{-b}$.
}
\label{tab1}
\end{table}
\begin{table}
\begin{center}
\begin{tabular}{llllll}
\hline
$\ell$&var&reg $\sqrt{r}$&reg $r$&non reg&non reg $V_G$\\[0.1 cm]
\hline
0&2.4[-9]&2.4[-9]&-7.6[-9]&6.9[-2]&6.9[-2]\\
1&1.7[-20]&1.6[-20]&2.5[-19]&-1.0[-3]&1.8[-20]\\
2&8.6[-7]&7.8[-7]&2.3[-6]&8.3[-7]&8.6[-7]\\
\hline
\end{tabular}
\end{center}
\caption{Relative error $\epsilon_{\rm rel}$ on the ground-state energy of the Coulomb potential $V(r)=-1/r$ for the $s$, $p$, and $d$ waves obtained from several Lagrange meshes (see text for details) with $N=10$ and $h=0.9$. The notation $a[-b]$ stands for $a \times 10^{-b}$.
}
\label{tab2}
\end{table}

In all cases, both regularized Lagrange meshes provide accurate results for the different partial waves and the $\sqrt{r}$-regularized Lagrange mesh provides energies closer to the variational calculation than the $r$-regularized Lagrange mesh.
Now, let us analyze the results in more details by starting with the harmonic-oscillator potential. 
For the $s$ wave, there is no centrifugal barrier and then, no singular term to be evaluated at the Gauss approximation. All meshes give thus similar accuracy. 
For the $p$ wave, the centrifugal matrix element is \textit{singular} for the non-regularized Lagrange mesh and for the $\sqrt{r}$-regularized Lagrange mesh. The non-regularized mesh is much less accurate if the centrifugal term is calculated at the Gauss approximation. However, in spite of the \textit{singularity}, the $\sqrt{r}$-regularized Lagrange mesh remains highly accurate.
For the $d$ wave, the presence of a \textit{singularity} coming from centrifugal term in the non-regularized mesh does not restrict anymore the accuracy of the method and all meshes are accurate. This is also true for higher partial waves.
The fact that the non-regularized Lagrange mesh is accurate for the $d$ wave has already been noticed in \cite{BHV02}.

Now, let us analyze the results obtained for the Coulomb potential. First, it should be pointed out that increasing the size of the model space by five units leads to a gain of five orders of magnitude in the accuracy for the $s$ and	 $d$ waves and even more for the $p$ wave. However, it is more instructive to restrict the model space to $N=10$ mesh points, which already provides sensible results. For specific values of the scaling parameter, $h=0.5$, $h=1$, or $h=1.5$, the Lagrange meshes can provide the exact lowest-state solutions of the Schr\"odinger equation for $\ell=0$, $\ell=1$, or $\ell=2$, respectively. These special cases, which could restrict the validity of our conclusions, are avoided by setting $h=0.9$. This value being close to the optimal scaling parameter of the lowest $p$ state, the results are more accurate for the $p$ wave than for the other. For the $s$ wave, the calculations based on the non-regularized Lagrange mesh are quite inaccurate due to the \textit{singularity} induced by the Coulomb potential. For the $p$ wave, the Coulomb \textit{singularity} is not problematic anymore since if the centrifugal term is calculated exactly, the non-regularized Lagrange mesh becomes as accurate as the variational calculation. Again, for the $d$ wave (and for the higher partial waves), all meshes are accurate.
For all partial waves, both regularized Lagrange meshes provide a similar accuracy as the variational calculation and the $\sqrt{r}$-regularized-Lagrange-mesh energies are closer to the variational energies than the $r$-regularized-Lagrange-mesh ones as in the harmonic-oscillator study.

In conclusion, the presence of \textit{singularities} in the matrix elements of the Hamiltonian between Lagrange functions leads sometimes to an important loss of accuracy and other times not. Let us try to understand why. The variational solution, denoted $\varphi_{\rm var}$, verifies \eqref{eigen}, which can be rewritten as
\begin{equation}\label{eigen2}
\langle f_i | H_\ell | \varphi_{\rm var}\rangle =E_{\rm var}  \langle f_i |\varphi_{\rm var}\rangle,
\end{equation}
where $E_{\rm var}$ designates the variational energy. 
The mesh solution will be close to the variational one if the matrix elements in \eqref{eigen2} are evaluated accurately at the Gauss approximation. 
This is true, in practice, if these matrix elements are not \textit{singular}, i.e.,\ if 
\begin{equation}
\frac{f_i(r) [H_\ell \varphi_{\rm var}(r)]}{w(r)}\ {\rm and}\ \frac{f_i(r) \varphi_{\rm var}(r)}{w(r)}
\end{equation}
are not singular, where $w$ is the weight function of the Gauss quadrature associated with the considered mesh. 
The second quotient is regular for all meshes and the first one presents a singularity, if any, only at the origin.  
However, when the variational solution $\varphi_{\rm var}$ is accurate, it behaves near the origin like the exact wave function, i.e., like $r^{\ell+1}$, \begin{equation}\label{q1}
\frac{f_i(r) [H_\ell \varphi_{\rm var}(r)]}{w(r)}
\mathop{\approx}\limits_{r\to 0} \frac{f_i(r) [H_\ell r^{\ell+1} \rme^{-r/2h}]}{w(r)},
\end{equation}
which differs, for $\ell \neq 0$, from the near-origin behavior of the Lagrange functions considered here,
\begin{equation}\label{q2}
\frac{f_i(r) [H_\ell f_j(r)]}{w(r)}\mathop{\approx}\limits_{r\to 0}  \frac{f_i(r) [H_\ell r \rme^{-r/2h}]}{w(r)}.
\end{equation}
As discussed before for the harmonic and Coulomb potentials, there is no clear link between the singularity of the quotients \eqref{q2} and the accuracy of the Lagrange-mesh results.
In contrast, the quotients \eqref{q1} are singular only when the non-regularized mesh is used for the $s$ wave in the case of the Coulomb potential and for the $p$ wave due to the centrifugal term in $H_\ell$.  These are precisely the only cases studied in Tables \ref{tab1} and \ref{tab2} where highly accurate energies are not obtained.

Following this simple reasoning, one expects that it is not the presence of \textit{singularities} in the matrix elements between Lagrange functions evaluated at the Gauss approximation which leads to a loss of accuracy but the presence of \textit{singularities} in the matrix elements between a Lagrange function and the variational solution.  The validity of this reasoning is also tested in the appendix by applying the Lagrange-mesh methods for solving two-dimensional radial Schr\"odinger equations.

Let us now consider the calculation of phase shifts. Since the simplicity of the Lagrange-mesh approaches comes from the systematic use of the Gauss approximation, only the regularized meshes, which have been proved to be accurate for all partial waves in the previous bound-state calculations, are considered here. 
As a first application, an Eckart potential is considered~\cite{Ec30},
\begin{equation}\label{VBa}
V(r)=-4b^2 \beta \frac{\rme^{-2br}}{(1+\beta \rme^{-2br})^2}
\end{equation}
with $\beta=(b-c)/(b+c)$.
The values $b=2$ and $c=-1$ are considered in the numerical applications. 
Since the $s$-wave phase shifts are known analytically~\cite{Ba49},
\begin{equation}\label{d0Ba}
\tan \delta_0(E)=\frac{\sqrt{2E}(b-c)}{2E+bc},
\end{equation}
only this partial wave is studied.

The s-wave phase shifts are evaluated by calculating first the pseudostates for either a $\sqrt{r}$-regularized Lagrange mesh with $\alpha=1$ or a $r$-regularized Lagrange mesh with $\alpha=0$. The number of mesh points is set to $N=15$ and the scaling parameter to $h=0.1$. Then, the phase shifts associated with the energies of the pseudostates are evaluated from \eqref{KVP} and \eqref{Ig}. The energies of the pseudostates can be varied by modifying the value of the scaling parameter $h$. The integrals in \eqref{KVP} and \eqref{Ig} are computed by means of the Gauss quadrature associated with the mesh, which makes their evaluation particularly easy.
The sensitivity of the results on the $\gamma$ parameter appearing in \eqref{KVP} is analyzed to determine the optimal range of $\gamma$-values, i.e. , the one leading to the minimal variation of the phase shift with respect to a variation of $\gamma$.
The $s$-wave phase shifts associated  with the energies of the first, the fifth, and the tenth pseudostates are given in Table~\ref{tab3}. 
\begin{table}
\begin{center}
\begin{tabular}{c r r r}
\hline
$\ell=0$&$E_1=0.1982139$&$E_{5}=4.95146$&$E_{10}=41.7$\\
\hline
reg $\sqrt{r}$&-49.67024& 50.0666&18.7\\
exact            &-49.67021& 50.0668&18.6\\[0.03 cm]
\hline
$\ell=0$&$E_1=0.2145073$&$E_{5}=5.38561$&$E_{10}=49.6$\\
\hline
reg $r$& -51.35794&48.3033&17.4\\
exact   & -51.35790&48.3036&17.1\\[0.03 cm]
\hline
\end{tabular}
\end{center}
\caption{$s$-wave phase shifts (in degrees) associated with the Eckart potential \eqref{VBa} with $b=2$ and $c=-1$. They are obtained by a $\sqrt{r}$-regularized-Lagrange-mesh calculation (reg $\sqrt{r}$) or a $r$-regularized-Lagrange-mesh calculation (reg $r$) from \eqref{KVP} and \eqref{Ig} with $N=15$ and $h=0.1$. The energies correspond to the ones of the first ($E_1$), the fifth ($E_5$), and the tenth ($E_{10}$) pseudostates. 
Exact $s$-wave phase shifts are evaluated from the analytic expression \eqref{d0Ba}.}
\label{tab3}
\end{table}
For each pseudostate, the phase shifts are stable around $\gamma=4$.
Highly accurate results are obtained with both Lagrange meshes. An error as small as about $4\times10^{-5}$ (resp. $3\times10^{-4}$) degree is obtained for the first (resp. fifth) pseudostate of the $s$ wave. The error for the tenth pseudostate is much bigger, of the order of $0.1$ degree but still impressive regarding the fact that the calculation is performed with only $15$ mesh points!

As a second example, I consider the $\alpha+\alpha$ potential proposed in \cite{BFW77}, which reproduces the $s$, $d$, and $g$ experimental $\alpha+\alpha$ phase shifts up to about $20$~MeV. It is defined by
\begin{equation}\label{VBu}
V(r)=V_0\ \rme^{-0.22 r^2}+4 \frac{e^2}{r} {\rm erf}(3 r/4),
\end{equation}
where ${\rm erf}$ designates the error function, $V_0=-122.6225\,{\rm MeV}$, and $e^2=1.44\,{\rm MeV\,fm}$. The calculations are performed with  $\hbar^2/M=20.736\,{\rm MeV\,fm}^2$.
The $s$- and $d$-wave phase shifts associated  with the energies of the first two pseudostates are given in Table~\ref{tab4}. 
\begin{table}
\begin{center}
\begin{tabular}{c r r }
\hline
$\ell=0$&$E_1=0.0105$&$E_{2}=1.8474$\\
\hline
reg $\sqrt{r}$&179.97 &116.67\\
"exact"          &179.96&116.63\\[0.03 cm]
\hline
$\ell=0$&$E_1=0.0107$&$E_{2}=1.9797$\\
\hline
reg $r$&179.96&112.64\\
"exact"&179.96&112.65\\[0.03 cm]
\hline
\hline
$\ell=2$&$E_1=2.10795$&$E_{2}=3.4183$\\
\hline
reg $\sqrt{r}$&12.471&94.460\\
"exact"          &12.470&94.464\\[0.03 cm]
\hline
$\ell=2$&$E_1=2.19462$&$E_{2}=3.5442$\\
\hline
reg $r$&15.123&99.596\\
"exact"          &15.120&99.600\\[0.03 cm]
\hline
\end{tabular}
\end{center}
\caption{$s$- and $d$-wave phase shifts (in degrees) associated with the  $\alpha+\alpha$ potential \eqref{VBu}. They are obtained by a $\sqrt{r}$-regularized-Lagrange-mesh calculation (reg $\sqrt{r}$) or a $r$-regularized-Lagrange-mesh calculation (reg $r$) from 
\eqref{KVP} and \eqref{Ig} with $N=15$ and $h=0.23$. The energies (in MeV) correspond to the ones of the two first ($E_1$ and $E_2$) pseudostates.}
\label{tab4}
\end{table}
They are obtained from a $\sqrt{r}$-regularized Lagrange mesh (reg $\sqrt{r}$) or a $r$-regularized Lagrange mesh (reg $r$) with $N=15$ and $h=0.23$.
To probe the method at both resonant and non-resonant energy, the value of $h$ is such as, for $\ell=2$, the energies of the two first pseudostates are, respectively, below and around the resonance energy.
The results are compared with the phase shifts obtained by a $R$-matrix calculation on a Lagrange-Legendre mesh~\cite{HSV98,Ba15} with large channel radius and large number of mesh points. They are referred as "exact" in Table~\ref{tab4} because they are expected to be numerically exact up to the number of digits quoted in this table. For the $s$-wave, for the first pseudostate, the analysis of the $\gamma$-sensitivity does not allow a clear distinction of a plateau. For consistency, the optimal $\gamma$ value obtained for the second pseudostate is thus used for both energies. 
The $\gamma$-sensitivity of the phase shifts is larger than with the Eckart potential. The optimal value of $\gamma$ depends on the pseudostate, the partial wave, and the mesh. For the cases considered in Table~\ref{tab4}, it varies between $0.3$ and $1.3$~${\rm fm}^{-1}$.   
In spite of the small number of mesh points, in all cases, highly accurate phase shifts are obtained.
\section{Conclusion}\label{S5}
A Lagrange-Laguerre mesh regularized by $\sqrt{r}$ is studied and compared with the non-regularized Lagrange-Laguerre mesh and the Lagrange-Laguerre mesh regularized by $r$. 
Contrary to the non-regularized mesh, both regularized meshes lead to accurate bound-state energies for singular potentials like the Coulomb and the centrifugal potentials, when the Gauss quadrature is used for evaluating the kinetic and potential matrix elements.
The analysis of these results has led to new insight about the effects of a \textit{singularity} on the accuracy of the Lagrange-mesh methods.

The $\sqrt{r}$-regularized Lagrange mesh, by contrast with the $r$-regularized Lagrange mesh, is based on \textit{exactly} orthonormal Lagrange functions. This property could be advantageous for an approximate variational approach based on Lagrange basis functions but where the Gauss quadrature will not be used for the computation of all matrix elements. Such an approach could be interesting for studying polyelectronic atoms as stated in the conclusion of \cite{BFG14}.

Finally, the Lagrange-mesh method based on both regularized Lagrange-Laguerre meshes has been proven to provide an highly accurate way to calculate phase shifts with small number of mesh points. The possibility to extend this approach to coupled-channel collisions or to the study of three-body phase shifts is worth investigating in the future.
\ack
I thank Daniel Baye for enlightening discussions about the Lagrange-mesh methods and his comments on the manuscript.
TRIUMF receives federal funding via a contribution agreement with the National Research Council Canada.
\appendix
\section*{Appendix: Two-dimensional Schr\"odinger equation}\label{A1}
\setcounter{section}{1}
In this appendix, I consider the Lagrange-mesh calculation of the bound-state solutions of a radial Schr\"odinger equation for a two-dimensional system,
\begin{equation}\label{Sch2}
\mathcal{H}_\ell u_{k m}(\rho)=\left(-\frac{1}{2}\frac{\rmd^2}{\rmd\rho^2} + \frac{4 m^2-1}{8 \rho^2}+V(\rho)\right) u_{k m}(\rho)=E u_{k m}(\rho),
\end{equation}
behaving at the origin like
\begin{equation}\label{ori2}
 u_{k m}(\rho) \mathop{\longrightarrow}\limits_{\rho\to 0} \rho^{m+1/2},
\end{equation}
where $\rho$ is the radius in polar coordinates.
Equation~\eqref{Sch2} is solved by expanding the radial wave function $u_{k m}$ on a $\sqrt{\rho}$-regularized Lagrange mesh $\{\tilde{f}^{(\alpha)}_j\}$ with $\alpha=0$. 
This mesh is able to reproduce the near-origin behavior \eqref{ori2} of the radial wave function for any partial wave, provided $N>m$. Let us note that although the matrix elements of $\rmd^2/\rmd\rho^2$ and $1/\rho^2$ between these Lagrange functions are divergent, the matrix elements of $\rmd^2/\rmd\rho^2+1/4\rho^2$ are convergent. They are given by
\begin{eqnarray}
\langle \tilde{f}^{(0)}_i |\frac{\rmd^2}{\rmd \rho^2}+\frac{1}{4 \rho^2}|\tilde{f}^{(0)}_j\rangle &=\langle \tilde{f}^{(0)}_i |\left(\frac{\rmd^2}{\rmd \rho^2}+\frac{1}{4 \rho^2}\right)\tilde{f}^{(0)}_j\rangle_G\\
&=(-1)^{i-j+1}  \frac{2}{(r_i-r_j)^2}
\end{eqnarray}
for $i\ne j$ and by 
\begin{eqnarray}
\langle \tilde{f}^{(0)}_i |\frac{\rmd^2}{\rmd \rho^2}+\frac{1}{4 \rho^2}|\tilde{f}^{(0)}_i\rangle
&=\langle \tilde{f}^{(0)}_i |\left(\frac{\rmd^2}{\rmd \rho^2}+\frac{1}{4 \rho^2}\right)\tilde{f}^{(0)}_i\rangle_G\\
&=\frac{-1}{12 r_i} \left[2(2N+1)-r_i-\frac{2}{r_i}\right]
\end{eqnarray}
 for $i=j$.
For $m\ne 0$, the matrix elements of $m/2 \rho^2$ diverge. However, if the Gauss quadrature is used, some finite values are obtained for these matrix elements and the Hamiltonian matrix can then be diagonalized without any specific issue. 

The accuracy of the Lagrange-mesh method is studied for the two-dimensional radial harmonic oscillator potential $V(\rho)=\rho^2/2$ and for the two-dimensional radial Coulomb potential $V(\rho)=-1/\rho$ for $m=1$. The results are displayed in Table~\ref{tab5}.
To evaluate accurately the relative errors, the calculations have been performed with quadruple-precision arithmetic.
\begin{table}
\begin{center}
\begin{tabular}{rrr}
\hline
$V(r)$&var&reg $\sqrt{\rho}$\\
\hline
$\rho^2/2$&3.0[-13]&2.1[-13]\\
$-1/\rho$&1.2[-16]&1.0[-16]\\
\hline
\end{tabular}
\end{center}
\caption{Relative error $\epsilon_{\rm rel}$ on the lowest-state energy of the two-dimensional radial potential $V(\rho)=\rho^2/2$ and $V(\rho)=-1/\rho$ for $m=1$ obtained from a $\sqrt{\rho}$-regularized Lagrange mesh and from the corresponding variational calculation. The calculations are performed with $N=20$ and $h=0.09$ for the harmonic potential and with $N=10$ and $h=0.9$ for the Coulomb potential. The notation $a[-b]$ stands for $a \times 10^{-b}$.
}
\label{tab5}
\end{table}
The number of mesh points $N$ and the value of the scale parameter $h$ are given in the table caption. The variational calculation is performed with a $\sqrt{\rho}$-regularized Lagrange basis of $N-1$ functions, with $\alpha=2$. 
Therefore, in both the mesh and the variational calculations, the radial wave function is expanded as a polynomial in $\rho$ of degree up to $N-1$ times $\sqrt{\rho}$ times an exponential.
The Lagrange-mesh method is as accurate as the variational approach while it involves the substitution of divergent integrals by their Gauss-quadrature approximation! 
Following the reasoning made in the analysis of the accuracy of the Lagrange-mesh method for the three-dimensional radial Schr\"odinger equation, this striking property could be anticipated. Indeed, the Gauss quadrature induces a huge loss of accuracy in presence of a singularity in the quotient $f_i(\rho) [\mathcal{H}_\ell \varphi_{\rm var}(\rho)/w(\rho)$, not in the quotient $f_i(\rho) [\mathcal{H}_\ell f_j(\rho)]/w(\rho)$. However, the quotient
\begin{equation}
\frac{f_i(\rho) [\mathcal{H}_\ell \varphi_{\rm var}(\rho)]}{w(\rho)}
\mathop{\approx}\limits_{\rho\to 0} \frac{f_i(\rho) [\mathcal{H}_\ell \rho^{\ell+1} \rme^{-\rho/2h}]}{w(\rho)}
\end{equation}	
is regular for all values of $m$.


\section*{References}
\begin{thebibliography}{11}
\bibitem{BH86}
Baye D and Heenen P-H 1986 \JPA {\bf 19} 2041--59

\bibitem{Ba15}
Baye D 2015 {\it Phys. Rep.} {\bf 565} 1--107

\bibitem{BHV02}
Baye D, Hesse M and Vincke M 2002 {\it Phys. Rev. }E {\bf 65} 026701

\bibitem{VMB93}
Vincke M, Malegat L and Baye D 1993 \jpb {\bf 26} 811-26

\bibitem{He02}
Hesse M 2002 {\it Phys. Rev. }E {\bf 65} 046703

\bibitem{KVB10}
Kievsky A, Viviani M, Barletta P, Romero-Redondo C and Garrido E 2010 {\it Phys. Rev. }C {\bf 81} 034002

\bibitem{Ec30}
Eckart C 1930 {\it Phys. Rev.} {\bf 35} 1303-09

\bibitem{Ba49}
Bargmann V 1949 {\it Rev. Mod. Phys.} {\bf 21} 488-93

\bibitem{BFW77}
Buck B, Friedrich H and Wheatley C 1977 {\it Nucl. Phys. }A {\bf 275} 246-68

\bibitem{HSV98}
Hesse M, Sparenberg J-M, Van Raemdonck F and Baye D 1998 {\it Nucl. Phys. }A {\bf 640} 37-51

\bibitem{BFG14}
Baye D, Filippin L and Godefroid M 2014 {\it Phys. Rev. }E {\bf 89} 043305

\end{thebibliography}
\end{document}